\documentclass[fleqn,usenatbib]{mnras}

\usepackage{newtxtext,newtxmath}
\usepackage[T1]{fontenc}

\DeclareRobustCommand{\VAN}[3]{#2}
\let\VANthebibliography\thebibliography
\def\thebibliography{\DeclareRobustCommand{\VAN}[3]{##3}\VANthebibliography}

\usepackage{color}
\definecolor{forestgreen}{rgb}{0.0, 0.5, 0.0}
\definecolor{chestnut}{rgb}{0.8, 0.36, 0.36}
\definecolor{byzantine}{rgb}{0.74, 0.2, 0.64}
\definecolor{orange}{rgb}{1.0,0.55,0.0}

\newcommand{\Mjup}{\mbox{M$_{\mathrm{jup}}$}}
\newcommand{\Msun}{\mbox{M$_{\odot}$}}
\newcommand{\Rsun}{\mbox{R$_{\odot}$}}

\newcommand{\Mwd}{\mbox{$M_{\mathrm{WD}}$}}

\newcommand{\lppr}{\stackrel{<}{\scriptstyle \sim}}
\newcommand{\lappr}{\raisebox{-0.4ex}{$\lppr$}}
\newcommand{\gppr}{\stackrel{>}{\scriptstyle \sim}}
\newcommand{\gappr}{\raisebox{-0.4ex}{$\gppr$}}

\usepackage{graphicx}	% Including figure files
\usepackage{amsmath}	% Advanced maths commands
\usepackage{gensymb}
\title[Origin of a close giant planet around a WD]{WD\,1856\,b: a close giant planet around a white dwarf that could have survived a common-envelope phase}

\author[F. Lagos et al.]{
F. Lagos,$^{1, 2, 3}$\thanks{E-mail: felipe.lagos@postgrado.uv.cl (FL)}
M. R. Schreiber,$^{4, 2}$
M. Zorotovic,$^{1}$
B. T. G\"ansicke,$^{5, 6}$
M. P. Ronco,$^{7, 2}$
Adrian S. Hamers$^{8}$
\\
$^{1}$ Instituto de F\'isica y Astronom\'ia, Universidad de Valpara\'iso, Avda. Gran Breta\~na 1111, Valpara\'iso, Chile\\
$^{2}$ N\'ucleo Milenio Formaci\'on Planetaria - NPF, Valpara\'iso, Chile \\
$^{3}$ European Southern Observatory (ESO), Alonso de Cordova 3107, Vitacura, Santiago, Chile\\
$^{4}$ Departamento de F\'isica, Universidad T\'ecnica Federico Santa Mar\'ia, Av. España 1680, Valpara\'iso, Chile \\
$^{5}$ Department of Physics, University of Warwick, Coventry CV4 7AL, UK\\
$^{6}$ Centre for Exoplanets and Habitability, University of Warwick, Coventry CV4 7AL, UK\\
$^{7}$ Instituto de Astrof\'isica, Pontificia Universidad Cat\'olica de Chile, Santiago, Chile.\\
$^{8}$Max-Planck-Institut f\"ur Astrophysik, Karl-Schwarzschild-Str. 1, 85741 Garching, Germany
}

\date{Accepted XXX. Received YYY; in original form ZZZ}

\pubyear{2015}

\begin{document}
\label{firstpage}
\pagerange{\pageref{firstpage}--\pageref{lastpage}}
\maketitle

\begin{abstract}
The discovery of a giant planet candidate orbiting the white dwarf WD\,1856+534 with an orbital period of 1.4\,d poses the questions of how the planet reached its current position.
We here reconstruct the evolutionary history of the system assuming common envelope evolution as the main mechanism that brought the planet to its current position. We find that common envelope evolution can explain the present configuration if it was initiated when the host star was on the AGB, the separation of the planet at the onset of mass transfer was in the range 1.69--2.35\,au, and if in addition to the orbital energy of the surviving planet either recombination energy stored in the envelope or another source of additional energy contributed to expelling the envelope. We also discuss the evolution of the planet prior to and following common envelope evolution.
Finally, we find that if the system formed through common envelope evolution, its total age is in agreement with its membership to the Galactic thin disc. We therefore conclude that common envelope evolution is at least as likely as alternative formation scenarios previously suggested such as planet-planet scattering or Kozai-Lidov oscillations.
\end{abstract}


%%%%%%%%%%%%%%%%%%%%%%%%%%%%%%%%%%%%%%%%%%%%%%%%%%

%%%%%%%%%%%%%%%%% BODY OF PAPER %%%%%%%%%%%%%%%%%%

\section{Introduction}

The vast majority of the stars in the Galaxy, including our Sun and virtually all known planet hosts will end their lives as white dwarfs. It is now well established that planets can survive the transformation of their host star into a white dwarf \citep{Villaver_2007,Mustill_2012,Rao_2018,roncoetal20}. Indirect observational evidence for the existence of planets around white dwarfs comes from the detection of photospheric metal absorption lines found in $20-30$\,per cent of all white dwarfs \citep{Zuckerman_2010, Koester_2014, Hollands_2017}, the discovery of dusty \citep{zuckerman+becklin87-1,farihietal09-1,Rocchetto_2015, Wilson_2019} and gaseous discs \citep{gaensickeetal_2006, gaensickeetal_2007,dennihyetal_2020,melisetal_2020} around a small sub-sample of these white dwarfs, the detection of a transiting planetesimal \citep[][]{Vanderburg_2015} in the process of disintegration \citep{Gaensicke_2016}, orbital motion in the gaseous disc around the white dwarf SDSS\,J1228+1040 which potentially indicates the presence of a planetary remnant \citep{Manser66}, and  finally an evaporating giant planet in close orbit around the young and hot white dwarf WD\,J0914+1914 \citep{boris_matthias19}. In all the above cases, the planetary material in close orbit around the white dwarf is usually assumed to have been scattered inward by gravitational interactions with a larger body, most likely a massive gas giant planet
\citep{nordhaus+spiegel13-1,mustill_2018}, or by the Eccentric Kozai-Lidov mechanism if the white dwarf is a member of a wide binary \citep{stephanetal17-1}.

Very recently, \citet{Vanderburgetal20} reported the discovery of a giant planet candidate, WD\,1856\,b, orbiting the cool and old white dwarf WD\,1856+534 (hereafter WD\,1856) with a short orbital period ($\simeq1.4$\,d). The authors discussed several evolutionary scenarios that could in principle explain the close orbit of the massive planet ($\leq14\Mjup$), including common envelope (CE) evolution, dynamical instability in a multiplanet system through the Kozai-Lidov (KL) mechanism (the white dwarf belongs to a visual triple star system) or planet-planet scattering events, both followed by tidal circularization, as well as other less likely scenarios such as instabilities due to galactic tides, close stellar encounters, close encounter scattering events between two planets near the periastron, 
tidal dissipation during the stellar giant phase, and partial tidal disruption. 
In line with the generally accepted interpretation for planetesimals and even Neptune-mass planets found in the proximity of white dwarfs, \citet{Vanderburgetal20} 
concluded that dynamical instabilities through close encounters with an additional object in the system, which excited large orbital eccentricities and subsequent tidal circularization, 
is the most reasonable explanation for the close orbit of WD\,1856\,b. 

Here we take a closer look at one scenario that has been considered unlikely by \citet{Vanderburgetal20}, i.e. CE evolution.  
We reconstruct the CE phase following the evolution of the progenitor of the white dwarf
and find that CE evolution is 
in fact a plausible scenario to explain the current orbital configuration 
of the system.
As in several close binary systems that most likely 
formed through CE evolution, only a small fraction of recombination energy is required to contribute to the envelope ejection process.

\section{Reviewing the arguments against common envelope evolution}

\citet{Vanderburgetal20} concluded that a CE evolution was highly unlikely to be the origin of the planet's current proximity to the white dwarf based on two arguments. The authors highlighted that the known white dwarf (and subdwarf) plus brown dwarf binaries collated by \citet{nelsonetal18} differ from WD\,1856 as all have short ($\lappr4$\,h) orbital periods and large ($\gappr50\,\Mjup$) masses, i.e. rather different from the much longer period and lower mass of the planet candidate in WD\,1856, providing an empirical argument that it might not have survived CE evolution. However, the sample listed by \citet{nelsonetal18} is small, and likely subject to significant selection effects. The majority of these systems were identified via the detection of either eclipses \citep{geieretal_2011} or emission lines from the irradiated companion \citep{maxted_2006}. Companions with longer orbital periods have a rapidly dropping probability of eclipsing the white dwarf, and their surfaces intercept less flux from the white dwarf resulting in weaker or absent emission lines. The latter is of course also true for companions orbiting cooler white dwarfs. We note here that GD\,1400 is a nearby ($\simeq46$\,pc) white dwarf with a long-period ($\simeq10$\,h, \citealt{burleighetal_2011}) brown dwarf companion identified initially as a weak infrared excess \citep{farihietal_2004}.

The much larger sample of post CE binary stars with low-mass stellar companions spans a wider range of orbital periods 
and both the CE efficiency as well as the  contributions from additional energy sources 
required to reconstruct their evolutionary paths can vary \citep{nebotetal11-1,zorotovicetal14a}.
The longer period of WD\,1856\,b might therefore, on its own, not represent a strong argument against CE evolution.  

\citet{Vanderburgetal20} also performed a simple reconstruction of CE evolution based on an analytic study of the parameterized equation that relates the binding energy of the envelope and the change in orbital energy: 
\begin{equation}
\label{eq:ce}
\frac{G M_\mathrm{1} M_\mathrm{e}}{\lambda R_\mathrm{L}} = \alpha_\mathrm{CE}\left[\frac{G M_\mathrm{c} M_\mathrm{2}}{2 a_\mathrm{f}}-\frac{G M_\mathrm{1} M_\mathrm{2}}{2 a_\mathrm{i}}\right]
\end{equation} 
where $M_\mathrm{1}$, $M_\mathrm{e}$, $M_\mathrm{c}$ and $M_\mathrm{2}$ are the masses of the primary star (the white dwarf progenitor), its envelope, its core (the future white dwarf) and the companion (the planet candidate), respectively, at the onset of CE evolution. $R_\mathrm{L}$ is the equivalent volume radius of the Roche-lobe (defined by $V=4/3\pi R_{\mathrm{L}}^3$ with $V$ being the Roche-volume), usually calculated with the Eggleton formula \citep{eggleton83-1}. At the onset of CE evolution this radius is identical to the radius of the primary star. $a_{\mathrm{i}}$ and $a_\mathrm{f}$ correspond to the orbital separation at the onset of the CE phase and immediately after the CE is ejected. The CE efficiency, $\alpha_\mathrm{CE}$, corresponds to the fraction of orbital energy lost during the CE phase that is used to unbind the envelope of the white dwarf progenitor, while $\lambda$ is a binding energy parameter that depends on the mass and evolutionary stage of the giant star. 
\citet{Vanderburgetal20} neglected the second term in the right-hand side of Eq.\,\ref{eq:ce} and used an analytical expression for the maximum radius of the star that only depends on its core mass (Eq.\,8 in their article). Using those assumptions, they derived the orbital period at the onset of the CE phase which combined with Eq.\,\ref{eq:ce} translates into an expression for the minimum $\alpha_\mathrm{CE}\lambda$ needed to unbind the envelope (avoiding a merger) that only depends on the measured parameters (white dwarf mass, companion mass and orbital period) and on the mass of the white dwarf progenitor.
\citet{Vanderburgetal20} found that the values of $\alpha_\mathrm{CE}\lambda$ they derived were larger than the tabulated results of \citet{xuli10-1}, which were computed for different progenitor masses and for a range of typical radii, even when including recombination energy, and concluded that CE evolution is a highly unlikely scenario.
However, a caveat to that conclusion is that the resolution in stellar mass of the tables from \citet{xuli10-1} is very coarse (1\Msun), and that these authors did not consider mass loss prior to the envelope ejection. It is evident from \citet[][their Fig.\,1]{xuli10-1} that the $\lambda$ parameter has a completely different behavior when changing from one to two solar masses, even reversing the slope of its dependence on the radius. A finer grid, especially in the range of $1-2$\,\Msun, which brackets the likely initial mass of the white dwarf progenitor of WD\,1856, is essential to derive a firmer conclusion. 

Here we therefore perform a more detailed reconstruction of the CE phase, considering mass loss prior to this phase, the possible contribution of recombination energy to the process of ejection of the envelope, and proper calculations for $\lambda$ depending on the mass and the evolutionary state of the primary star.

\section{Reconstructing common envelope evolution}
\label{sec:CE} 

\begin{figure*}
\centering
 %\begin{subfigure}
  \includegraphics[width=\columnwidth]{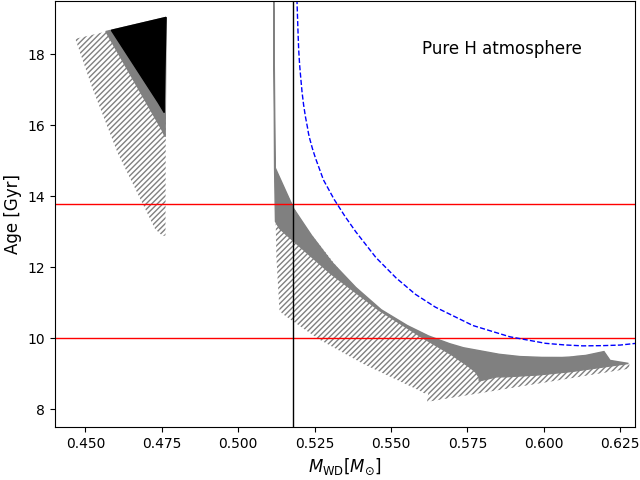}
 %\end{subfigure}
 \hfill
 %\begin{subfigure}
  \includegraphics[width=\columnwidth]{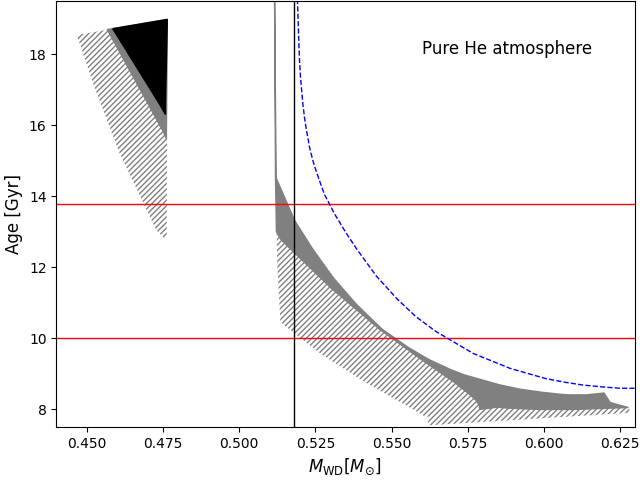}
 %\end{subfigure}
\caption{Total system age of WD\,1856 as a function of the white dwarf mass derived from our reconstruction for CE evolution and adopting cooling ages for pure-hydrogen (left panel) and pure-helium (right panel) atmospheres. We assumed that recombination energy contributed not at all to the ejection of the CE (solid black area), less than 10~per~cent (solid gray area), or up to 100~per~cent (shaded gray area). For comparison, the blue dashed line corresponds to the estimated age of the white dwarf assuming single star evolution (i.e. no CE phase). We searched for possible CE progenitor systems across a range of white dwarf masses in the range $0.408-0.628$, i.e. bracketing the best-fit measurement (0.518\,\Msun, vertical black line, \citealt{Vanderburgetal20}) by twice the uncertainty of the measurement ($\pm0.055\,\Msun$). The upper and lower horizontal red lines correspond to the age of the Universe ($13.8$~Gyr) and the estimated age of the oldest stars within the thin disc of the Galaxy ($10$~Gyr), respectively. CE solutions exist for white dwarf masses in the considered white dwarf mass range and with total ages below the age of the thin disk while single star evolutionary models require a slightly large white dwarf mass. }
\label{fig:CE_sim}
\end{figure*}

Based on its kinematics, \citet{Vanderburgetal20} found that WD\,1856 most likely belongs to the thin disc of the Galaxy, which puts an upper limit of $\simeq10$\,Gyr on the age of the system. 
As already noted by \citet{Vanderburgetal20}, if the white dwarf formed from single star evolution, this age limit is in conflict with the best-fit white dwarf mass ($0.518\pm0.055\,\Msun$), i.e. the sum of the white dwarf cooling age and the main-sequence life time of its progenitor exceed the age of the thin disc.

We revisit this problem by calculating the total age of WD\,1856 using the single star evolution (SSE) code \citep{hurleyetal2000} for the main sequence and post main sequence evolution and the white dwarf models of \citet{bedardetal20} to determine the cooling age of the white dwarf. Given the atmosphere of WD\,1856 is most likely a mixture of hydrogen and helium \citep{Vanderburgetal20}, we computed the age for the two limiting cases, where, for a given mass, the pure hydrogen and pure helium case represent an upper and lower limit on the cooling age of the white dwarf.
The total age of WD\,1856, if formed through single star evolution, is illustrated by the blue dashed line in Fig.\,\ref{fig:CE_sim}, clearly indicating that the white dwarf mass has to be $\ga0.57\,\Msun$ for its age to be below 10\,Gyr, with the exact value depending on the (uncertain) atmosphere composition.
If the white dwarf mass is as small as measured by \citet{Vanderburgetal20}, every evolutionary scenario  for WD\,1856 and its close-in planet that assumes the white dwarf has formed through single star evolution, implies an age of the white dwarf in conflict with its membership to the thin disk of the Galaxy. CE evolution can truncate the core-growth of a giant star and therefore more massive stars, with a shorter main sequence lifetime, can be the progenitors of low-mass white dwarfs, which 
reduces the total age of the system.  

To evaluate the possibility of CE evolution for
WD\,1856 we here use an algorithm that we previously applied to
reconstruct CE evolution of white dwarfs with close
low-mass star companions 
\citep{zorotovicetal2010,zorotovicetal14a}. 
In brief, the algorithms works as follows. 
We calculated a grid of
stellar evolution tracks 
using SSE with initial masses up to eight solar
masses (with a step size of $0.01$\Msun). 
As CE evolution truncates the core growth of the giant star, 
the core mass of the giant star at the onset of CE evolution 
must have been equal to the white dwarf mass in the post-CE system. 
We therefore first search our grid for giant stars with a core mass equal to the measured white dwarf
mass. As we know the radius and the mass of each of the identified potential
white dwarf progenitor stars, we can for all of them determine the orbital period at
the onset of CE evolution using the mass of the companion
and Roche-geometry. In other words,  
we can determine the period and stellar masses
for every possible progenitor system. 
Knowing the stellar masses and the period at the onset of CE evolution and comparing it with those of the post-CE system, provides the change in orbital energy during CE evolution
for each of these potential progenitor binary star systems.
As we also know the 
amount of recombination energy stored in the envelope of the giant at the onset of CE evolution,
computed as in \citet{claeysetal_2014} and \citet{hurleyetal2002},
we can test whether sufficient energy is
available to unbind the envelope of the giant. If this is the case, we
consider this particular system a possible progenitor configuration.
Searching our finely space gird,
we can therefore find virtually all possible progenitors 
for a given white dwarf mass, post-CE orbital period, and companion mass. 
If the effective temperature of the white dwarf is known as well, we
can also calculate the total age of each potential progenitor system
by summing the cooling age of the white dwarf 
(obtained from the tracks of \citealt{bedardetal20})
as well as the main sequence and post
main sequence lifetime of the progenitor star (obtained from SSE). Our reconstruction algorithm 
does not consider the evolution of the system prior to mass transfer. We will discuss possible evolutionary 
scenarios that lead to the onset of CE evolution in the next section.

We searched for CE progenitor configurations for WD\,1856  over a range of white dwarf masses covering twice the uncertainty in mass provided by \citet{Vanderburgetal20}, i.e. $0.408-0.628\,\Msun$. 
We used the measured current orbital period of $1.41$\,d, and adopted the largest possible mass of $14\,\Mjup$ for the planet, which corresponds to the upper end of the mass range derived by \citet{Vanderburgetal20}, and again the two limiting cases for the composition of the atmosphere (pure hydrogen and pure helium). 
The shaded regions in Fig.\,\ref{fig:CE_sim} show the possible progenitor systems that we identified as a function of total age and  white dwarf mass. 
We distinguish three cases for the contribution of recombination energy to unbind the CE: no recombination energy used (i.e. $\alpha_\mathrm{rec}=0$, black area), less than 10~per~cent of contribution (i.e. $0<\alpha_\mathrm{rec}\leq0.1$, solid gray area), and up to 100~per~cent of contribution (i.e. $0<\alpha_\mathrm{rec}\leq1$, shaded gray area). 
The vertical line shows the best-fit white dwarf mass of \citet{Vanderburgetal20}, and the two lower and upper horizontal red lines give the estimated age for the thin disc of the Galaxy and the Hubble time, respectively. 

To what degree recombination energy can contribute to CE ejection is intensively discussed. 
While \citet{ivanovaetal15-1} suggest that recombination energy can play a fundamental role, others 
\citep[e.g.][]{soker+harpaz2003} argue that the opacity of the envelope is too small for recombination energy to provide an important contribution. Based on the latter argument and on the fact that the evolutionary history 
of two post-CE binary star systems with long orbital periods (IK\,Peg and KOI-3278) can be understood 
by assuming a small fraction (1 and 2 per cent) of recombination energy to contribute \citep{zorotovicetal2010,zorotovicetal14a}, we here consider as realistic those cases in which the efficiency of recombination energy remains below 10 per cent. 
Table\,\ref{tab:CE} lists the orbital parameters, initial and at the onset of CE evolution, for the possible solutions, taking into account the restriction in age (total age below $10$\,Gyr) and keeping the fraction of recombination energy below 10~per~cent (dark grey and black regions in Fig.\,\ref{fig:CE_sim}). The values provided for $\lambda$ include contributions from recombination energy. Therefore $\lambda$ can become significantly larger than one \citep[see also e.g.][]{dewi+tauris2000}
and the left-hand side of Eq.\,\ref{eq:ce} corresponds to the binding energy reduced by a fraction of recombination energy contributing to unbinding the envelope.

All possible progenitors that filled their Roche-lobes on the first giant branch ($\Mwd < 0.5 \Msun$ in Fig.\,\ref{fig:CE_sim}) exceed the age limit significantly, reaching total ages even larger than the Hubble time for smaller white dwarf masses. This includes post-CE binaries in which the white dwarf evolves first into a Helium burning horizontal branch star and the envelope is fully expelled.
In all the reconstructed cases that fulfill the age (below $10$\,Gyr) and recombination energy constraint (not more than $10$ per cent), the progenitor had to be on the thermally pulsating asymptotic giant branch (TPAGB) phase when it filled its Roche lobe. During this phase the binding energy parameter $\lambda$ can take values well above unity when recombination energy is considered \citep[see Fig.~5 in][]{camachoetal14}, which implies that the envelope can be very loosely bound. Therefore, even considering only a small fraction of the available recombination energy to contribute to the ejection process allows us to reconcile the current configuration of the white dwarf and the surrounding planet candidate with CE evolution. 
Also, according to our reconstruction code, the progenitor of the white dwarf must have already lost $\simeq20$~per~cent of its initial mass when the CE phase occurred, which facilitates the ejection. 

\begin{table}
	%\centering
	\caption{Range of reconstructed parameters for the white dwarf progenitor + planet candidate binary assuming a CE evolution, restricting $\alpha_\mathrm{rec}\leq0.1$ and the total age of the system to be less than 10\,Gyr.}
	\resizebox{\columnwidth}{!}{\begin{tabular}{llccc}
	\hline
	Parameter & Pure-H model & Pure-He model\\
	\hline
	Initial mass of the progenitor (\Msun) & $1.51 - 2.31$ & $1.41 - 2.31$ \\
    Mass of the progenitor at CE onset (\Msun) & $1.20 - 2.07$ &  $1.09 - 2.07$ \\
	Age of the progenitor at CE onset (Gyr) & $1.00 - 3.03$ &  $1.00 - 3.78$ \\
    Orbital separation at CE onset (au) & $1.69 - 2.35$  & $1.69 - 2.35$ \\
    Radius of the progenitor at CE onset (au) & $1.23 - 1.73$ & $1.23 - 1.73$ \\
	Post-CE white dwarf mass (\Msun) & $0.561 - 0.628$ & $0.551 - 0.628$ \\
    $\alpha_\mathrm{CE}$ & $0.36 - 1.00$ & $0.36 - 1.00$ \\
    $\alpha_\mathrm{rec}$ & $0.04 - 0.10$ & $0.04 - 0.10$ \\
    $\lambda$ & $3.72 - 10.95$ & $3.08 - 10.95$ \\
    \hline
	\end{tabular}}
	\label{tab:CE}
\end{table}

The large number of possible solutions we find with our reconstruction algorithm clearly illustrates that CE evolution represents a plausible scenario for the formation of WD\,1856 and its close planetary companion. In particular, CE evolution allows for a larger range of masses and younger ages for the white dwarf than formation scenarios that assume the white dwarf formed through single star evolution. 

However, it is important to keep in mind that so far we assumed a planetary mass corresponding to the upper limit provided by \citet{Vanderburgetal20}, i.e. $14$\,\Mjup. If the mass of the planet is smaller than this upper limit, which cannot be excluded,
the number of possible solutions found by our reconstruction algorithm decreases. We ran our algorithm for smaller planet masses and
for values below five \Mjup\, we did not find solutions with ten per cent or less of the available recombination energy contributing to expelling the envelope. If the planet mass can be measured in the future and turns out to be below five \Mjup, 
CE evolution should therefore be considered a rather unlikely scenario. For larger planet masses, CE evolution represents a convincing explanation for the existence of the system.

\section{Prior to common envelope evolution}
\label{sec:migra} 

In order to trigger CE evolution, the planet WD\,1856\,b must have been located at $1.69-2.35$\,au at the end of the main sequence evolution of its host star. This separation can be achieved early during the pre-main sequence due to angular momentum exchange between the planet and the protoplanetary disc. The classical core accretion formation scenario \citep{Mizuno_1980,BodenheimerPollack_1986,Pollack_1996}, along with type~I and II disc migration \citep{Ward_1997,Tanaka_2002,LinPapaloizou_1986,Armitage_2007}, can naturally lead to formation of planets like WD\,1856\,b at locations in the range of $\sim1-10$\,au \citep[see][for details]{Mordasini_2018}. 
Furthermore, even planets formed through gravitational instability can be subject to disc migration \citep{Boss_2005, ChaNayakshin_2011}. 
In particular, with 2D hydrodynamical simulations \citet{Baruteau_2011} showed that these planets can migrate inward due to their interactions with the disc on the fast type~I migration-like time-scales. 

If the planet around WD\,1856 migrated inward to separations of a few au while the protoplanetary disk was still present, it 
must later have survived the first giant branch evolution of the host star and triggered mass transfer and a subsequent CE phase. 
We tested how realistic this scenario is using the code by \citet{roncoetal20} but applying the eddy turnover time scale  
given in \citet{villaver+livio09-1} in the description of tidal forces.  
We found that for an initial stellar mass of 1.5M$_\odot$ and considering stellar tides, the initial semi-major axis for a 10$M_{\text{jup}}$ planet to survive the RGB and to possibly generate CE evolution during the AGB is beyond $2.20$\,au.  For larger stellar masses (e.g $2$M$_{\odot}$), smaller separations are possible as the tip of the first giant branch is very short and therefore tides during the first giant branch have less impact on the planet separation. It is therefore entirely possible that the planet migrated inward to a separation exceeding a few au, survived the first giant branch evolution, was brought further in due to tides when the star was on the AGB, and finally triggered mass transfer when reaching a separation in the range of $1.69-2.35$\,au as predicted by our reconstruction of CE evolution.

\begin{figure}
    %\hspace{-2em}% 
    \includegraphics[width=\columnwidth]{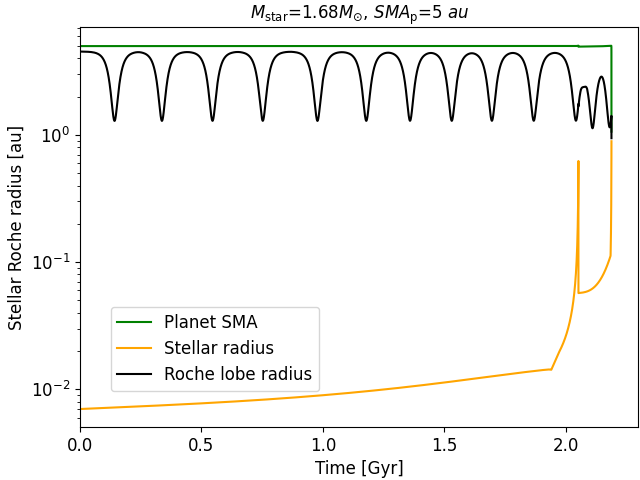}%\hspace{-2em}% 
  \caption{Secular evolution of the stellar Roche lobe radius for a $1.68\, \Msun$ white progenitor being orbited by a $14\, \Mjup$ planet with a semi-major axis of 5 au.
  The gravitational perturbation exerted by the M-dwarf binary located at 1000 au trigger KL oscillations on the planet, which combined with tidal forces during the red and asymptotic giant branch approaches the planet to the star until the latter fills its Roche lobe during the asymptotic giant branch. The initial eccentricities for the planet, M-dwarf binary and outer orbit are 0.001, 0.3 and 0.3 respectively. Both the argument of periapsis ($\omega$) and longitude of the ascending node ($\Omega$) are set to 0 for the outer and M-dwarf binary orbits. For the planet's orbit we set $\omega_\mathrm{p}=60\degree$ and $\Omega_\mathrm{p}=80\degree$.}
  \label{fig:1}          
\end{figure}

Given that WD\,1856 and its planetary companion are members of a multiple star system, with a distant M-dwarf binary, an alternative evolution of the system prior to CE evolution are KL oscillations \citep{kozai_1962,Lidov_1962}.
In order to evaluate this possibility, we used the \textsc{SecularMultiple} 
code \citep{hamersetal2016,hamersetal2018,hamersetal2020}, which allows to calculate the secular orbit-averaged evolution of the planet during the main sequence evolution of the host star, 
including the effect of tidal bulges, tidal friction and general relativity. 
We also implement the single-star-evolution (\textsc{SSE}) code \citep{hurleyetal2000} to model radial expansion, contraction, structure changes, and mass loss due to stellar evolution of the white dwarf progenitor. 
We found that for the case of 1.41 and 2.31\,\Msun\ white dwarf progenitor masses and initial semi-major axis exceeding 4\,au, 
the star can experience Roche lobe overflow during the asymptotic giant branch as required to trigger CE evolution. While the exact outcome of the quadruple dynamics is very sensitive on 
the initial values of the argument of periapsis, longitude of ascending nodes, and eccentricity of the outer binary, 
and in particular the mutual inclination,  
this shows that quadruple dynamics offer an alternative mechanism that might have brought WD\,1856\,b close enough to generate a CE phase. 
Figure \ref{fig:1} shows one of the simulations in which Roche-lobe overflow and therefore CE evolution is triggered (black line).   
This simulation also illustrate that the assumption for initially circular planetary orbits in our reconstruction algorithm 
is reasonable as tidal forces circularize the orbit before dynamically unstable mass transfer leads to CE evolution. 

%%%%%%%%%%%%%%%%%%%%%%%%%%%%%%%%%%%%%%%%%%%%%%%%%%

\section{Evaporation during and after the common envelope phase}

While stellar companions clearly survive orbiting within a CE (as long as they do not merge with the core of the giant star), this is not necessarily true for planetary mass companions. In fact, \citet{nelemans+tauris} considered evaporating or merging planetary 
mass companions as a possible explanation for observed single low-mass white dwarfs without a stellar companion. In order to evaluate whether the planet candidate around WD\,1856 would have survived inside the envelope, we here follow their approach. 

As the planet will move inside the envelope during CE evolution, it can be completely evaporated. 
\citet{soker_1998} suggested that this occurs if the local sound speed ($c_{\rm s}$) in the envelope reaches the escape velocity ($v_{\rm esc}$) of the planet, i.e. 
\begin{equation}
  c_{\rm s}^2 \approx v_{\rm esc}^2 \Longleftrightarrow
   \gamma\,\frac{k_{\rm B}T}{\mu \,m_{\rm u}} \approx
   \frac{2\,G\,M_{\rm p}}{R_{\rm p}},
\end{equation} 
where $R_{\mathrm{P}}$ and $M_{\mathrm{P}}$ are the radius and the mass of the planet, and $T$ is the temperature of the envelope at the distance from the core ($r$). 
Using this 
criterion and assuming the same temperature structure of the envelope as \citet{nelemans+tauris}, i.e.    
\begin{equation}
    T \simeq 1.78 \times 10^6 \times (r/\mathrm{R_\odot})^{-0.85} \mathrm{K},  
\end{equation}
we can estimate at which separation form the core (the future white dwarf) a planet with a 
given mass and radius would be fully evaporated inside the giant stars envelope. 
For the measured parameters of WD\,1856 we find that for separations exceeding $0.71\Rsun$ the 
planet should survive inside the giants envelope which is well below the current separation of the planet.  

As has been shown by \citet{bear-soker_2011}, evaporation may remain an issue for planets 
that survived CE evolution.  
Indeed, immediately after the envelope has been expelled, the strong EUV radiation by the 
hot white dwarf generates hydrodynamic escape in hydrogen dominated atmospheres up to orbital distances 
of several astronomical units and may even generate detectable accretion onto the hot white dwarf \citep{Schreiber_2019}. 
To estimate whether WD\,1856\,b could have survived this phase of strong EUV evaporation at its current position, we use simple scaling laws for hydrodynamic escape.  

\begin{figure}
    \includegraphics[width=1.0\columnwidth]{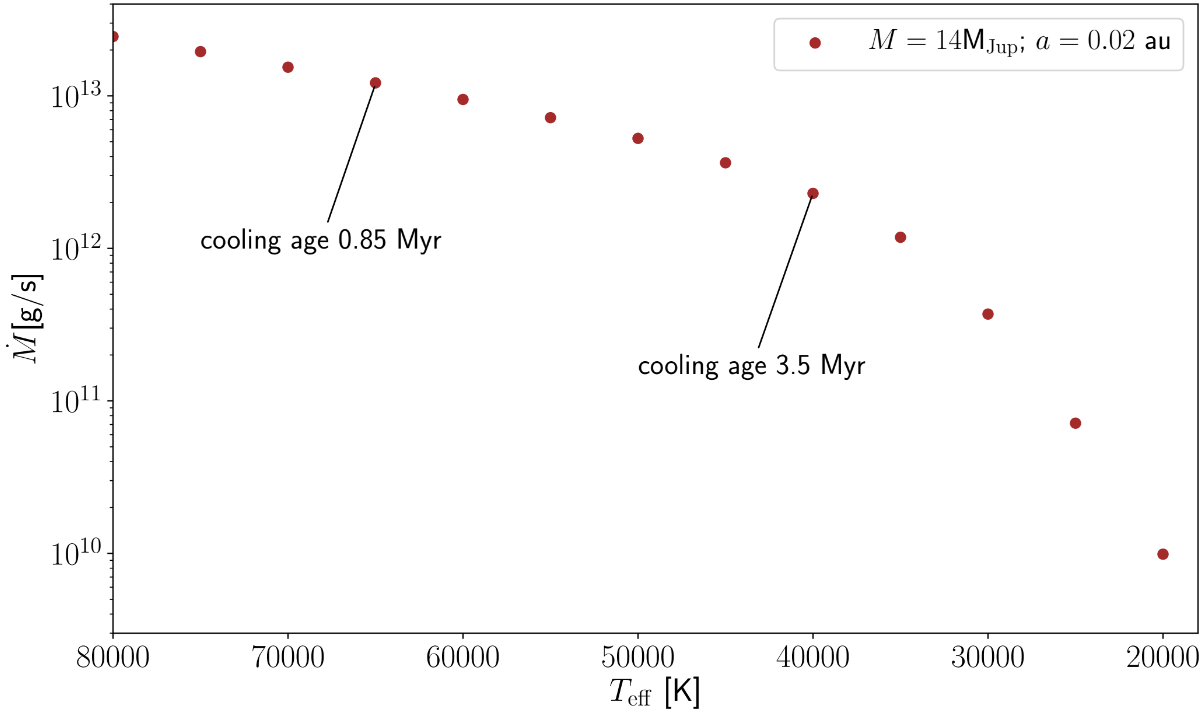}  
  \caption{Mass loss rate of the planet due to the strong EUV irradiation by the hot white dwarf following CE evolution as a function of effective temperature. While the mass loss rate must have led to detectable accretion for several million years, the total mass loss is far below one per cent of the planets mass and is clearly negligible.}
  \label{fig:3}          
\end{figure}

For large EUV fluxes, mass loss due to hydrodynamic escape can be in the energy limited 
or the recombination limited regime. In the first case mass loss is proportional to the EUV irradiation
while it scales with the square root of the EUV irradiation in the latter. 
For a Jupiter-mass planet, the transition between both regimes is usually assumed to occur at $10\,000\,\mathrm{erg\,cm^{-2}s^{-1}}$. 
For the mass loss rate in the energy limited regime of irradiated giant planets we used \citep{Murray-Clay_2009}: 
\begin{equation}
    \dot{M}=\frac{\epsilon \pi F_{\mathrm{EUV}} R^{3}_{\mathrm{P}}}{GM_{\mathrm{P}}K(\xi)} 
\end{equation}
where $F_{\mathrm{EUV}}$ is the incident EUV flux, and $\epsilon$ the efficiency of using the incident energy, which we set to $\epsilon=0.3$ \citep{Murray-Clay_2009,Owen-alvarez_2016}. 
As WD\,1856\,b is quite close to the white dwarf, where the Roche-lobe ($R_{\mathrm{L_P}}$) and planet radius become somewhat comparable, mass loss is slightly enhanced. This is accounted for by the correction term $K(\xi=R_\mathrm{L_P}/R_\mathrm{P})$ for which we used equation 17 from \citet{Erkaev_2007}. 
We calculated the EUV fluxes from the same model spectra used in \citet{Schreiber_2019,boris_matthias19} and used the above equations to estimate the mass loss rate. 
Figure\,\ref{fig:3} shows the resulting mass loss rate as a function of effective temperature (and cooling age). 
The total mass loss is clearly negligible and we therefore conclude that neither during nor after CE evolution the planet was significantly affected by evaporation. 

\section{Concluding discussion}
\label{sec:discussion}

We have re-evaluated CE evolution as a possible mechanism to explain the 
existence of the gas giant planet candidate in a close 1.4\,day orbit around the white dwarf 
WD\,1856. 
Reconstructing the CE evolution we found that it offers a reasonable evolutionary scenario for this system. 
There exist plenty of mechanisms, including triple dynamics, that could have brought the planet to the location required to trigger CE evolution (separations of $1.69-2.35$\,au). 
The planet would also not have suffered important mass loss due to evaporation, neither during nor after CE evolution. 
Most importantly, we find that during CE evolution enough energy is available to unbind the envelope. 
If only $4-10$~per cent of the available recombination energy contributed to expelling the envelope and the mass of the planet is between five\,\Mjup\, and the upper limit of 14\,\Mjup\, derived by \citet{Vanderburgetal20}, the evolution of WD\,1856 can be fully understood in the context of CE evolution, similar to the cases of the post-CE binary stars IK\,Peg and KOI-3278 \citep{zorotovicetal14a}.

Our results clearly show that if energy in addition to the orbital energy provided by the spiral-in of the planet contributed to expelling the envelope, CE evolution can explain the current configuration of WD\,1856. In our simulations we assumed this additional energy to be recombination energy. However, in theoretical considerations of CE evolution the opinions on whether recombination energy can indeed significantly contribute to the envelope ejection process differ drastically. 
While \citet{Webbink_2008} as well as more recently \citet{ivanovaetal15-1} and \citet{ivanova18-1} found that recombination energy can play an 
important role, others argue that most of the recombination energy is radiated away and that it therefore has a limited impact on the envelope ejection process \citep[e.g.][]{sabachetal17-1,gricheneretal18-1,sokeretal18-1}. 
If recombination energy turns out to be indeed largely inefficient, other energy sources might have helped the planet to unbind the envelope of the giant. Examples for possible energy sources that have been discussed, but are usually considered lees likely to provide important contributions, include accretion energy, nuclear energy, enthalpy \citep[see][for more details]{ivanova_2013}, or jets \citep{sabachetal17-1}.
Recently, \citet{wilson+nordhaus19-1} added that the ejection efficiency is sensitive to the properties of the surface-contact convective region and that including convection may aid ejection without the need for additional energy sources.
Potentially most interesting in the case of WD\,1856, however, is the possibility that orbital energy of an additional giant planet played a role. This hypothetical additional planet could have initially been located somewhat closer to the star than WD\,1856\,b and therefore did not survive the metamorphosis of the host star, but a 
fraction of its orbital energy could have helped WD\,1856\,b in its effort to expel the giant envelope.
A scenario involving inner planets that are engulfed has been suggested earlier by \citet{bear+soker11-1} to explain the existence of the later refuted planet around the horizontal branch star HIP\,13044.
As this alleged planet had a long orbital period (16.2\,days) for a post CE system, the theoretical scenarios that were developed towards explaining its existence are worth to be considered for WD\,1856\,b. Apart from the multi-planet suggestion, 
the incomplete envelope ejection scenario proposed by \citet{bearetal11-1} could apply to WD\,1856\,b. According to this model, the planet was engulfed for a short period of time at the end of the first giant branch evolution of the host star and only a fraction of the common envelope was ejected 
\citep[see also][]{passyetal12-1}.

Given the large variety of evolutionary scenarios involving a CE, given that WD\,1856 is not alone with its need for additional energy, and given the fact that several realistic mechanisms 
that might have provided this extra energy are in principal available, CE evolution clearly represents 
a reasonable scenario for the evolutionary history of WD\,1856 and its close planetary companion. 

Alternative scenarios that have been suggested for the formation of WD\,1856 and its close planetary companion are 
planet-planet scattering \citep{Vanderburgetal20} and triple dynamics 
\citep{oconnor2020,munoz2020,stephanetal20-1}. 
An important observational constraint on possible evolutionary scenarios would be provided by a mass measurement of the planet. The currently available observations allow for planet masses between 14\,\Mjup\, and $0.1$\,\Mjup. The eccentric Kozai-Lidov mechanism works only for planet masses as small as $1-3\,\Mjup$ \citep{munoz2020}
and also planet-planet scattering is more likely to bring lower mass planets to close orbits around the white dwarf \citep{Maldonado2020}. 
In contrast, CE evolution predicts the planet mass to most likely exceed five $\Mjup$.

However, for all alternative scenarios to CE evolution, the likely membership of WD\,1856 to the Galactic thin disc, limiting its age to $\simeq10$\,Gyr, poses an additional problem if the white dwarf mass is as small as measured by \citet{Vanderburgetal20}.
This conclusion does not depend on the assumed metallicity of the white dwarf progenitor or the initial to final mass ratio. 
Table\,\ref{tab:ts} lists the white dwarf progenitor masses for a sample of initial to final mass ratios and the corresponding main sequence lifetime derived using SSE, assuming a white dwarf mass of 0.52\Msun. The time since the end of the main sequence until the progenitor becomes a white dwarf is around 1-1.7\,Gyr, depending on the assumed metallicity. The cooling age, calculated from the \citet{bedardetal20} cooling sequences, lies between 5.77 and 6.11\,Gry depending on the composition of the white dwarf atmosphere. Adding the post main
sequence lifetime and the cooling age to the main sequence lifetimes listed in Table\,\ref{tab:ts}, we find that all of the listed initial to final mass ratios lead to total ages exceeding the age of the Galactic thin disc (10\,Gyr), even assuming the shortest times derived for the post main sequence evolution and the white dwarf cooling. We therefore conclude that CE evolution is the most likely scenario for the history of WD\,1856 if the white dwarf mass is $\simeq0.52\pm0.05\Msun$ as measured by \citet{Vanderburgetal20}. If the white dwarf mass is somewhat larger, i.e. $\simeq0.6\Msun$, CE evolution remains one possibility together with alternative models such as close encounters with other planets in the system or Kozai-Lidov oscillations triggered by the distant companions. As CE evolution is the only scenario currently available consistently solving the age issue, we predict the planet mass to be above five $\Mjup$.

\begin{table}
    \centering
    \caption{Main-sequence mass and lifetime for a 0.52\,\Msun\ white dwarf progenitor. The progenitor masses were derived using five different initial to final mass relations from the literature (first five rows) and the evolutionary code SSE \citep{hurleyetal2000} with three different metallicities (last three rows). The main-sequence lifetimes were all derived using SSE, assuming solar metallicity for the first five rows.}
    \begin{tabular}{lll}
    \hline
     $\mathrm{M_{ms}}$($\Msun$) & $\tau_\mathrm{ms}$(Gyr) & IFMR reference \\
    \hline
     1.154                  & 6.45            & \citet{Weidemann_2000} \\
     1.403                  & 3.35            & \citet{Williams_2009} \\
     0.948                  & 13.53             & \citet{catalan_2008} \\
     1.363                  & 3.66            & \citet{Casewell_2009}\\
     1.156                  & 6.65            & \citet{Kalirai_2008} \\
     0.885                  & 14.51           & SSE (z = 0.01) \\
     1.020                  &  10.20           & SSE (z = 0.02) \\
     1.170                  &  6.83           & SSE (z = 0.03) \\
     \hline
    \end{tabular}
    \label{tab:ts}
\end{table}

%%%%%%%%%%%%%%%%%%%%%%%%%%%%%%%%%%%%%%%%%%%%%%%%%%

%\section{Conclusions}
%
%CE evolution offers realistic scenario for the formation of ...

%%%%%%%%%%%%%%%%%%%%%%%%%%%%%%%%%%%%%%%%%%%%%%%%%%

\section*{Acknowledgements}
We thank the referee Noam Soker for helpful comments and suggestions and acknowledge fruitful discussions with Cristobal Petrovich and Dimitri Veras.
FL is supported by the ESO studentship. MRS acknowledges support from FONDECYT Grant No.\,1181404.
MZ is supported by CONICYT PAI (Concurso Nacional de Inserci\'on en la Academia 2017, Folio 79170121) and CONICYT/FONDECYT (Programa de Iniciaci\'on, Folio 11170559). 
BTG was supported by a Leverhulme Research Fellowship and the UK STFC grant ST/T000406/1.
MPR acknowledges financial support provided by FONDECYT grant 3190336. 
FL, MRS, and MPR thank for support from the Iniciativa Cient\'{\i}fica Milenio (ICM) via the N\'ucleo Milenio de Formaci\'on Planetaria Grant.
ASH thanks the Max Planck Society for 
support through a Max Planck Research Group.
%%%%%%%%%%%%%%%%%%%%%%%%%%%%%%%%%%%%%%%%%%%%%%%%%%
\section*{Data Availability}

The data used in this article can be obtained upon request to Felipe Lagos (felipe.lagos@postgrado.uv.cl) and after agreeing to the terms of use.

%%%%%%%%%%%%%%%%%%%% REFERENCES %%%%%%%%%%%%%%%%%%% 
\bibliographystyle{mnras}
\bibliography{paper} % if your bibtex file is called paper.bib
%%%%%%%%%%%%%%%%%%%%%%%%%%%%%%%%%%%%%%%%%%%%%%%%%%

%%%%%%%%%%%%%%%%% APPENDICES %%%%%%%%%%%%%%%%%%%%%
%\appendix
%If you want to present additional material which would interrupt the flow of the main paper,it can be placed in an Appendix which appears after the list of references.
%%%%%%%%%%%%%%%%%%%%%%%%%%%%%%%%%%%%%%%%%%%%%%%%%%

% Don't change these lines
\bsp	% typesetting comment
\label{lastpage}
\end{document}